\documentclass[sigconf]{acmart}

\pdfoutput=1

\usepackage{booktabs} % For formal tables
\usepackage{tablefootnote}
\usepackage{balance}
\let\proglang=\textsf

\begin{document}

\copyrightyear{} 
\acmYear{} 
\setcopyright{none=true}
\settopmatter{printacmref=false, printccs=false, printfolios=false}

%\setcopyright{}
\acmConference[ICSE-NIER'18]{40th International Conference on Software Engineering: New Ideas and Emerging Results Track}{May 27-June 3, 2018}{Gothenburg, Sweden}
%\acmBooktitle{ICSE-NIER'18: 40th International Conference on Software Engineering: New Ideas and Emerging Results Track, May 27-June 3, 2018, Gothenburg, Sweden}
\acmPrice{}
\acmDOI{}
\acmISBN{}

\title{Replication studies considered harmful}

\author{Martin Shepperd}
\orcid{0003-1874-6145}
\affiliation{%
  \institution{Brunel University London}
  \streetaddress{Department of Computer Science, Kingston Lane}
  \city{Uxbridge} 
  \postcode{UB8 3PH}
}
\email{martin.shepperd@brunel.ac.uk}

\renewcommand{\shortauthors}{M Shepperd}

\begin{abstract}
\emph{\textbf{Context}}: There is growing interest in establishing software engineering as an evidence-based discipline.  To that end, replication is often used to gain confidence in empirical findings, as opposed to reproduction where the goal is showing the correctness, or validity of the published results. \newline
\emph{\textbf{Objective}}:  To consider what is required for a replication study to confirm the original experiment and apply this understanding in software engineering. \newline
\emph{\textbf{Method}}: Simulation is used to demonstrate why the prediction interval for confirmation can be surprisingly wide.  This analysis is applied to three recent replications.\newline
\emph{\textbf{Results}}: It is shown that because the prediction intervals are wide, almost all replications are confirmatory, so in that sense there is no `replication crisis', however, the contributions to knowledge are negligible.  \newline
\emph{\textbf{Conclusion}}: Replicating empirical software engineering experiments, particularly if they are under-powered or under-reported, is a waste of scientific resources.  By contrast, meta-analysis is strongly advocated so that all relevant experiments are combined to estimate the population effect.
\end{abstract}


 
\maketitle

\section{Introduction}
The idea of replicating empirical studies in order to enhance the trustworthiness of an empirical result is widely seen as a central tenet of the scientific method \cite{Open15}.  This paper addresses the rather overlooked questions---certainly within empirical software engineering---of \emph{(i) how do we know if a replication study confirms the original study} and \emph{(ii) were it to do so, what would this tell us?}  Likewise if the replication fails to confirm the original study, what can we learn from this?

Using empirical evidence, typically through experiment, observation and case study, to underpin software engineering has been gaining  traction in recent years. This has been stimulated in part by the seminal paper promoting evidence-based software engineering from Kitchenham et al.~\cite{Kitc04}.  Clearly it's desirable to understand which methods and techniques `work', to what extent, and in what contexts?  From this, emerged the idea of the community building knowledge through replication studies \cite{Shul08}.  A mapping study by da Silva et al.~\cite{daSi12} and extended by Bezerra et al.~\cite{Beze15} found 135 articles reporting a total of 184 replications (1994--2012).

However, recently concerns have been expressed about the reliability of empirical findings both within software engineering \cite{Jorg16} and beyond, e.g., in experimental psychology \cite{Open15}.  Consequently, replication studies have been seen as  important for two reasons.  First, in terms of their ability to increase our confidence in specific empirical findings via confirmation, or otherwise. Second, as a form of sample to estimate the reliability of software engineering empirical studies in general.

The remainder of this paper, briefly reviews the state of replication studies in software engineering focusing on experimentation.  Next I show, by simulation, that simply through sampling error we can obtain considerably more diverse results than might be imagined.  This is applied to a selection of published replication studies by formally computing prediction intervals. Finally I discuss the implications and argue that research effort would be far more usefully deployed performing meta-analyses.

\section{Related work}
A key work that sets out the generally accepted view of role of replication\footnote{N.B., my focus is on `conceptual' as opposed to `exact' replications that deal with reproducibility questions and, in any case, can be problematic when using human participants.}  studies in software engineering is by Shull et al.~\cite{Shul08}.  They state that:
\begin{quote}
``if the results of the original study are reproduced using different experimental procedures, \emph{then the community has a very high degree of confidence that the result is real}" \cite{Shul08} [my italics].  
\end{quote}
\noindent
It would be fair to say that this represents the majority view in empirical software engineering and the paper is highly cited\footnote{According to Google Scholar there are in excess of 200 citations (8.2.2018).}.  In other words, the primary purpose of replication is to increase confidence.  

But how similar must results be to constitute confirmation?  Curiously this has not been directly addressed, so whilst researchers generally feel able to make judgements concerning a replication, I am unaware of any replication study in software engineering that has stated in some statistical sense how close a result must be to constitute a confirmation.

A range of approaches have been deployed to make comparisons.  These include: comparison of (i) simple descriptors, e.g., means, (ii) goodness of fit measures, e.g., R-squared, (iii) correlations, (iv) null hypothesis significance testing (NHST) where both the original and replication study report statistical significance at an agreed $\alpha$ threshold and (v) standardised effect sizes such as Cohen's $d$ or Cliff's $\delta$.  Of these, NHST is the dominant paradigm.  

Unfortunately, NHST has come in for extensive criticism \cite{Cumm08,Colq14}.  For example, it has been argued that given the flexibility in choice of data and analysis methods the desire to have `positive' findings is likely to substantially increase the likelihood of a false-positive above the nominal level set by $\alpha$ typically 5\% \cite{Simm11,Gelm14}.  Another difficulty arising from the `all or nothing' nature of NHST is publication bias due to the preference of authors, reviewers and editors for `positive' results and the file-drawer problem \cite{Rose79}.  Examples are reported from psychology \cite{Masi12} and software engineering \cite{Jorg16} of the surprising prevalence of significant $p$ values.  Again this selectivity makes it more likely that a replication will fail to such a large effect as the original study \cite{Jorg16}.  

A further problem is experimental power.  For any experimental design, the power depends on sample size, measurement error, the number of comparisons being performed, and the effect size under investigation \cite{Gelm14}.  However, it is clear we work in a field that is dominated by low power studies \cite{Kamp07,Jorg16} and this is problematic in that it does not just mean a reduced likelihood of detecting true effects, it also implies increased likelihood of over-estimating the effect size or finding an effect that does not really exist \cite{Butt13}.  Finally, NHST is impacted by sample size, so if the original and replication studies employ different numbers of experimental units this alone might lead to different values of $p$. 

To summarise, although not generally made explicit, empirical software engineering researchers usually require both studies to report statistical significance (in the same direction) for the replication to be considered confirmatory.  The meaning of neither study being significant is less self-evident \cite{Colq14}.  This leaves a decision as to whether a replication `confirms' the original study as largely being a subjective judgement.

\section{Simulating replications} \label{Sec:Sim}
In order to give insight into the major role that sampling errors play on the variability of the experimental results, I use a simple Monte Carlo simulation\footnote{The \proglang{R} code, additional figures and associated materials are available from \url{https://figshare.com/articles/_/5873754}.}. Suppose we have two treatments X and Y and we want to compare them experimentally.  Each experiment has 30\footnote{J{\o}rgensen et al.~\cite{Jorg16} report that in their survey of software engineering experiments 47\% had a sample size of 25 or less.} units, where a unit might be a participant, a data set, and so forth.  Let's also suppose the experimental design is extremely simple and that the two samples are independent, as opposed to paired.  We also assume the rather unlikely situation of no measurement errors, no publication bias and file drawer problems \cite{Simm11} and also that the underlying population is normally distributed.

Starting with the simplest case of no effect with $\mu=100, \sigma=20$, 10000 simulated experimental results behave as might expected from the Central Limit Theorem.  However, as confirmed by Table~\ref{Tab:NoEffectDist} we actually observe a surprisingly wide range of possible effect sizes \cite{Cohe92} with only just over half the experiments finding negligible or no effect [-0.2, +0.2].  Note that these simulation circumstances are considerably more propitious than we might expect in real life \cite{Jorg16}.  

%\begin{figure}[htp]
%\begin{center}
%\caption{Histogram of Random Experiments with No True Effect}
%\label{Fig:NoEffectdHist}
%\includegraphics[width=\linewidth]{NoEffectdHist.png}
%\end{center}
%\end{figure}

\begin{figure}[htp]
\begin{center}
\caption{Boxplots of Random Experiments With and Without a True Small Effect}
\label{Fig:MCboxplots}
\includegraphics[height=5cm,keepaspectratio]{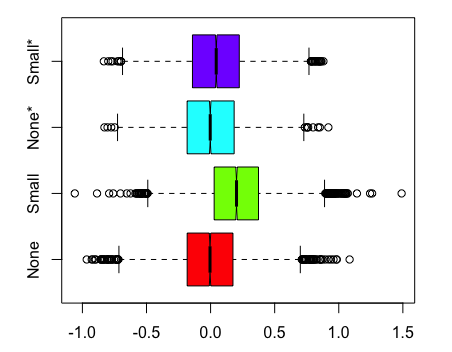}
\end{center}
\end{figure}

Next we contrast this simulation with a small positive effect using normal and rather more realistic mixed-normal distributions.  These contaminated or mixed normal distributions generate heavy tails that differ from strictly normal distributions due to the presence of additional outliers \cite[Chap.~1]{Wilc12}.  Figure \ref{Fig:MCboxplots} shows the boxplots of the experimental estimates of the effect size for these four cases (where None and Small are the true effect sizes and * denotes a distribution with outliers).  It is clear that there is (i) a great deal of variability in all the results and (ii) small departures from normality greatly hinder our ability to detect a true small effect, exemplified by the very similar distributions for Small* and None*.  %In an environment of selective publication (i.e., picking results from the extreme positive tail) it's easy to envisage the impact on the false discovery rate.

\begin{table}[htp]
\caption{Proportions of Different Effect Sizes (Using Cohen's d) for No True Effect}
\begin{center}
\begin{tabular}{|p{0.75cm}|p{0.7cm}|p{0.7cm}|p{0.7cm}|p{0.7cm}|p{0.7cm}|p{0.7cm}|p{0.7cm}|}
\hline
True&Large & Med  & Small & None & Small  & Med  & Large \\
 Effect &-ve & -ve & -ve & & +ve & +ve & +ve \\
 \hline
None & 0.1\% & 2.7\% & 19.5\% & 55.5\% & 19.7\% & 2.4\%  & 0.20\% \\
Small &   0.02\% & 0.27\% & 5.79\% & 43.61\% & 37.78\% & 11.19\% &  1.34\% \\
None* &   0.05\% & 2.54\% & 20.0\% & 54.62\% & 20.17\% & 2.57\% &  0.05\% \\
Small* &  0.04\% & 1.76\% & 16.71\% & 52.11\% & 25.08\% & 4.25\% & 0.05\% \\
\hline
\end{tabular}
\end{center}
\label{Tab:NoEffectDist}
\end{table}%

Finally, we simulate the replication process by randomly drawing pairs of studies, without replacement, and observing the difference in results.  Thus for each simulation of 10,000 experiments this yields 5,000 replications.  Table \ref{Tab:SimEffType} summarises the result in terms of confirmation, or Gelman's S-errors \cite{Gelm14}.  Unsurprisingly, when there is no true effect, the distribution of effect sign agreement is uniform.  In the event of a normal distribution, and no other nuisance factors, something like 60\% of replications will find an effect in the same direction as the original study, but not necessarily with much concordance in terms of effect size.  However, the presence of even a few outliers as per the mixed normal distributions reduces the number of replications that agree in direction to under one third.  And this simulation simply focuses on sampling error.  Introducing other sources of error e.g., measurement or excess researcher degrees of freedom \cite{Loke17} compound the situation.

 \begin{table}[htp]
\caption{Counts of Simulated Replication Effect Direction Agreement}
\begin{center}
\begin{tabular}{|l|r|r|r|r|}
\hline
True Effect &  - -
 &  - + &  + - &  + + \\
 \hline
None & 1249 & 1246  & 1250  & 1255 \\
Small &  244 &   825  &  855  & 3076 \\
None* & 1241  & 1268  & 1246  & 1245 \\
Small* & 938  & 1201 &  1234  & 1627 \\
 \hline
\end{tabular}
\end{center}
\label{Tab:SimEffType}
\end{table}%

\noindent
Thus we can see that even in quite benign circumstances, replication is likely to be quite hit or miss simply because of sampling errors.

\section{Confirmation of results in software engineering}

Next we consider the situation of replication studies within software engineering.  As previously noted, two mapping studies \cite{daSi12,Beze15} have located well in excess of 100 replications, however, unfortunately careful examination reveals few report any measures of dispersion, e.g., variance of the measure of effect or response variable.  This inhibits calculation of prediction intervals.  Table \ref{Tab:ExamplePredInts} shows three examples that are selected because some calculations are possible.  

As stated, replication studies in software engineering have not been in the habit of stating what range of values might be expected from a confirmatory replication.  Formally we are asking what variation might arise just from sampling error.  The Monte Carlo simulations from Section \ref{Sec:Sim} indicate that this source of variability can be surprisingly large.  The variability can be characterised as a prediction interval.  

In this analysis the 95\% prediction interval is reported using the approach due to Spence and Stanley \cite{Spen16} and implemented by the \proglang{R} package \texttt{predictionInterval}.  Note a prediction interval differs from a confidence interval because we are concerned with the estimate from the specific study being replicated, as opposed to an estimate of the population effect size.  Generally prediction intervals will be a little wider than confidence limits \cite{Stan14}.

\begin{table}[htp]
\caption{Example Replications and Associated Prediction Intervals}
\begin{center}
\begin{tabular}{|p{0.75cm}|p{0.75cm}|p{1.1cm}|p{1.6cm}|p{1cm}|p{1.2cm}|}
\hline
Orig Study & Rep Study & Orig Eff Sz & Pred Int & Rep Eff Sz & Confirms?\\
\hline
\cite{Shep97} & \cite{Myrt99} & 0.101 & [-0.33, 0.53]\tablefootnote{The interval is calculated by reasoning backwards from the replication study to the original study.} & -0.1\tablefootnote{The effect size is approximate due to the estimated pooled standard deviation.} & Y \\
\cite{Jorg03} & \cite{Shep05} & -0.176 & [-0.84, 0.48] & 0.122 & Y \\
\cite{Bria97} & \cite{Bria01} & 1.430 & [0.05, 2.76] & 1.090 & Y\\
\hline
\end{tabular}
\end{center}
\label{Tab:ExamplePredInts}
\end{table}%

What is particularly noteworthy in Table \ref{Tab:ExamplePredInts} are the wide prediction intervals so, for instance in the second example, experiment \cite{Shep05} would confirm the original experiment \cite{Jorg03} if anything from a large negative effect to a small-medium positive effect were detected.  There are two contributory factors. First, small sample sizes, e.g., \cite{Bria97}.  Second, small effect sizes, e.g., \cite{Shep97,Jorg03} which are often driven down by high variance e.g., \cite{Myrt99}.  Thus, hugely varying results can simply be explained by sampling error.  Of course this will in all probability be exacerbated by measurement error and publication bias so the foregoing prediction interval might be regarded as the best case scenario.

An alternative view is to regard studies that seek to answer the same research question as inputs to meta-analysis.  To illustrate this, a simple meta-analysis is undertaken using the standardised mean difference effect size approach of Lipsey and Wilson \cite{Lips01} and implemented in the \proglang{R} package of \texttt{Metafor} from Viechtbauer \cite{Viec10}.  The experimental results are pooled in order to estimate the population effect size.

\begin{figure}[htp]
\begin{center}
\includegraphics[width=\linewidth]{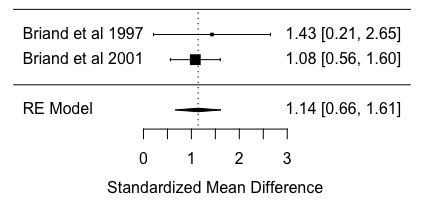}
\caption{Forest plot of combining studies through meta-analysis}
\label{Fig:ForestPlots}
\end{center}
\end{figure}

Fig.~\ref{Fig:ForestPlots} shows a forest plot of the two studies of the effect of good design on comprehension for OO systems from \cite{Bria97} and \cite{Bria01}.  The horizontal bars show the confidence intervals for the estimated effect sizes and the sizes of the centre points are proportional to the sample sizes.  The standardised mean distance is a measure of mean difference normalised by standard deviation.  Assuming a simple fixed effects model we obtain the an estimate of Cohen's d=1.14 [0.66, 1.61] denoted by the vertical dashed line.  Note the limits are at least all in one direction and are narrower than either study individually (see Table \ref{Tab:ExamplePredInts}).  Thus we gain knowledge and precision as opposed to simply reporting that we can confirm the original weak finding.

\section{Discussion and Conclusions}

The title of this paper is intentionally provocative.  The purpose, however, is to draw attention to the twin issues of how similar must a replication be to the original experiment to constitute confirmation and how effective is the process of replication for adding empirically-derived, software engineering knowledge?  This paper has only considered experiments but in parallel, there is a case for developing, and applying, strong meta-analytic methods for qualitative studies e.g., case studies and action research.

Clearly, there is a place for us to consider reproducibility.  That a study is reproducible should be considered minimally necessary \cite{Peng11}.  But beyond this, we have shown there are two distinct difficulties with replication studies as practised in software engineering.  First, when the prediction interval can be properly constructed \cite{Cumm08,Spen16} what constitutes a confirmation is often a good deal broader than might be anticipated.  This --- as has been shown both by simulation and by example --- could include a wide range of effect sizes, sometimes in \emph{both} directions. Thus confirmation, particularly for low-powered experiments and small effect sizes, can be trivial. 

In contrast, combining studies through meta-analysis enables all relevant studies to be combined so that we may generate our best estimate  (complete with confidence interval) of the effect in question. This yields more nuanced information than reducing the matter to an all or nothing matter of confirmation or disconfirmation.  Of course, meta-analyses cannot overcome the problems of poor quality primary studies or selective reporting and publication.  However, techniques such as funnel plots can at least help highlight these problems \cite{Schm14}.  Lack of heterogeneity (perhaps due to methodological differences or the existence of meaningful sub-populations) can also be detected and investigated \cite{Lips01,Schm14}. 

This implies the following recommendations for the empirical software engineering community:
\begin{enumerate}
\item Properly report studies and in particular provide information on the \emph{dispersion}, e.g., variance, of the response (dependent) variables.  Without this information neither the prediction interval can be computed (for replication analysis) nor is meta-analysis possible.  Given the current paucity of such information, this is the biggest single contributor to wasted research effort.
\item Construct prediction intervals \emph{prior} to conducting replication studies and understand that under-powered studies of small effects (i.e., much of empirical software engineering \cite{Kamp07,Jorg16}) can be trivially replicated, but the contribution to knowledge will be extremely small.
\item Limit replications to matters of reproducibility (where warranted).
\item Conduct, independent studies of important research questions where the effects may matter to practising software engineers and combine results using meta-analytic techniques.   Avoid close replications since these may violate the independency assumption underlying meta-analysis \cite{Kitc08}. Also, consider corrections to the meta-analysis \cite{Schm14} needed due to potential bias from inflated effect size estimates from the first study arising from publication bias \cite{Ioan08}.
\end{enumerate}
 
%``Given that replication studies will be more accepted in the future, one could imagine that it is quite easy to grab ``low-hanging fruit'' by replicating existing studies. Which degree of replication is healthy?" \cite{Sieg15} i.e., an efficient way to increase publication counts.

\begin{acks}
This work was partly funded by the EPSRC Grant EP/P025196/1.
\end{acks}

\balance
\bibliographystyle{ACM-Reference-Format}
\bibliography{NIER2018} 


\begin{thebibliography}{31}

%%% the Web address.

\ifx \showCODEN    \undefined \def \showCODEN     #1{\unskip}     \fi
\ifx \showDOI      \undefined \def \showDOI       #1{#1}\fi
\ifx \showISBNx    \undefined \def \showISBNx     #1{\unskip}     \fi
\ifx \showISBNxiii \undefined \def \showISBNxiii  #1{\unskip}     \fi
\ifx \showISSN     \undefined \def \showISSN      #1{\unskip}     \fi
\ifx \showLCCN     \undefined \def \showLCCN      #1{\unskip}     \fi
\ifx \shownote     \undefined \def \shownote      #1{#1}          \fi
\ifx \showarticletitle \undefined \def \showarticletitle #1{#1}   \fi
\ifx \showURL      \undefined \def \showURL       {\relax}        \fi
% invisible to TeX
\providecommand\bibfield[2]{#2}
\providecommand\bibinfo[2]{#2}
\providecommand\natexlab[1]{#1}
\providecommand\showeprint[2][]{arXiv:#2}

\bibitem[\protect\citeauthoryear{Bezerra, da~Silva, Santana, de~Magalh{\~a}es,
  and Santos}{Bezerra et~al\mbox{.}}{2015}]%
        {Beze15}
\bibfield{author}{\bibinfo{person}{R. Bezerra}, \bibinfo{person}{F. da Silva},
  \bibinfo{person}{A. Santana}, \bibinfo{person}{C. de Magalh{\~a}es}, {and}
  \bibinfo{person}{R. Santos}.} \bibinfo{year}{2015}\natexlab{}.
\newblock \showarticletitle{Replication of Empirical Studies in Software
  Engineering: An Update of a Systematic Mapping Study}. In
  \bibinfo{booktitle}{{\em Intl.\ Symp. on Emp.\ Softw.\ Eng.\ and
  Measurement}}. \bibinfo{pages}{1--4}.
\newblock


\bibitem[\protect\citeauthoryear{Briand, Bunse, and Daly}{Briand
  et~al\mbox{.}}{2001}]%
        {Bria01}
\bibfield{author}{\bibinfo{person}{L. Briand}, \bibinfo{person}{C. Bunse},
  {and} \bibinfo{person}{J. Daly}.} \bibinfo{year}{2001}\natexlab{}.
\newblock \showarticletitle{A controlled experiment for evaluating quality
  guidelines on the maintainability of object-oriented designs}.
\newblock \bibinfo{journal}{{\em IEEE Transactions on Software Engineering\/}}
  \bibinfo{volume}{27}, \bibinfo{number}{6} (\bibinfo{year}{2001}),
  \bibinfo{pages}{513--530}.
\newblock


\bibitem[\protect\citeauthoryear{Briand, Bunse, Daly, and Differding}{Briand
  et~al\mbox{.}}{1997}]%
        {Bria97}
\bibfield{author}{\bibinfo{person}{L. Briand}, \bibinfo{person}{C. Bunse},
  \bibinfo{person}{J.. Daly}, {and} \bibinfo{person}{C. Differding}.}
  \bibinfo{year}{1997}\natexlab{}.
\newblock \showarticletitle{An Experimental Comparison of the Maintainability
  of Object-Oriented and Structured Design Documents}.
\newblock \bibinfo{journal}{{\em Emp.\ Softw.\ Eng.\/}} \bibinfo{volume}{2},
  \bibinfo{number}{3} (\bibinfo{year}{1997}), \bibinfo{pages}{291--312}.
\newblock


\bibitem[\protect\citeauthoryear{Button, Ioannidis, Mokrysz, Nosek, Robinson,
  and Munaf{\`o}}{Button et~al\mbox{.}}{2013}]%
        {Butt13}
\bibfield{author}{\bibinfo{person}{K. Button}, \bibinfo{person}{J. Ioannidis},
  \bibinfo{person}{C. Mokrysz}, \bibinfo{person}{J. Nosek, B.and~Flint},
  \bibinfo{person}{E. Robinson}, {and} \bibinfo{person}{M. Munaf{\`o}}.}
  \bibinfo{year}{2013}\natexlab{}.
\newblock \showarticletitle{Power failure: why small sample size undermines the
  reliability of neuroscience}.
\newblock \bibinfo{journal}{{\em Nature Reviews Neuroscience\/}}
  \bibinfo{volume}{14}, \bibinfo{number}{5} (\bibinfo{year}{2013}),
  \bibinfo{pages}{365--376}.
\newblock


\bibitem[\protect\citeauthoryear{Cohen}{Cohen}{1992}]%
        {Cohe92}
\bibfield{author}{\bibinfo{person}{J. Cohen}.} \bibinfo{year}{1992}\natexlab{}.
\newblock \showarticletitle{A power primer}.
\newblock \bibinfo{journal}{{\em Pyschological Bulletin\/}}
  \bibinfo{volume}{112}, \bibinfo{number}{1} (\bibinfo{year}{1992}),
  \bibinfo{pages}{155--159}.
\newblock


\bibitem[\protect\citeauthoryear{Colquhoun}{Colquhoun}{2014}]%
        {Colq14}
\bibfield{author}{\bibinfo{person}{D. Colquhoun}.}
  \bibinfo{year}{2014}\natexlab{}.
\newblock \showarticletitle{An investigation of the false discovery rate and
  the misinterpretation of $p$-values}.
\newblock \bibinfo{journal}{{\em Royal Society Open Science\/}}
  \bibinfo{volume}{1}, \bibinfo{number}{3} (\bibinfo{year}{2014}).
\newblock


\bibitem[\protect\citeauthoryear{Cumming}{Cumming}{2008}]%
        {Cumm08}
\bibfield{author}{\bibinfo{person}{G. Cumming}.}
  \bibinfo{year}{2008}\natexlab{}.
\newblock \showarticletitle{Replication and $p$ intervals: $p$ values predict
  the future only vaguely, but confidence intervals do much better}.
\newblock \bibinfo{journal}{{\em Perspectives on Psychological Science\/}}
  \bibinfo{volume}{3}, \bibinfo{number}{4} (\bibinfo{year}{2008}),
  \bibinfo{pages}{286--300}.
\newblock


\bibitem[\protect\citeauthoryear{da~Silva, Suassuna, Fran\c{c}a, Grubb,
  Gouveia, Monteiro, and dos Santos}{da~Silva et~al\mbox{.}}{2012}]%
        {daSi12}
\bibfield{author}{\bibinfo{person}{F. da Silva}, \bibinfo{person}{M. Suassuna},
  \bibinfo{person}{A. Fran\c{c}a}, \bibinfo{person}{A. Grubb},
  \bibinfo{person}{T. Gouveia}, \bibinfo{person}{C. Monteiro}, {and}
  \bibinfo{person}{I. dos Santos}.} \bibinfo{year}{2012}\natexlab{}.
\newblock \showarticletitle{Replication of empirical studies in software
  engineering research: a systematic mapping study}.
\newblock \bibinfo{journal}{{\em Emp.\ Softw.\ Eng.\/}} \bibinfo{volume}{19},
  \bibinfo{number}{3} (\bibinfo{year}{2012}), \bibinfo{pages}{501--557}.
\newblock


\bibitem[\protect\citeauthoryear{Gelman and Carlin}{Gelman and Carlin}{2014}]%
        {Gelm14}
\bibfield{author}{\bibinfo{person}{A. Gelman} {and} \bibinfo{person}{J.
  Carlin}.} \bibinfo{year}{2014}\natexlab{}.
\newblock \showarticletitle{Beyond power calculations: Assessing Type S (sign)
  and Type M (magnitude) errors}.
\newblock \bibinfo{journal}{{\em Perspectives on Psychological Science\/}}
  \bibinfo{volume}{9}, \bibinfo{number}{6} (\bibinfo{year}{2014}),
  \bibinfo{pages}{641--651}.
\newblock


\bibitem[\protect\citeauthoryear{Ioannidis}{Ioannidis}{2008}]%
        {Ioan08}
\bibfield{author}{\bibinfo{person}{J. Ioannidis}.}
  \bibinfo{year}{2008}\natexlab{}.
\newblock \showarticletitle{Why most discovered true associations are
  inflated}.
\newblock \bibinfo{journal}{{\em Epidemiology\/}} \bibinfo{volume}{19},
  \bibinfo{number}{5} (\bibinfo{year}{2008}), \bibinfo{pages}{640--648}.
\newblock


\bibitem[\protect\citeauthoryear{J{\o}rgensen, Dyb{\aa}, Liest{\o}l, and
  Sj{\o}berg}{J{\o}rgensen et~al\mbox{.}}{2016}]%
        {Jorg16}
\bibfield{author}{\bibinfo{person}{M. J{\o}rgensen}, \bibinfo{person}{T.
  Dyb{\aa}}, \bibinfo{person}{K. Liest{\o}l}, {and} \bibinfo{person}{D.
  Sj{\o}berg}.} \bibinfo{year}{2016}\natexlab{}.
\newblock \showarticletitle{Incorrect Results in Software Engineering
  Experiments: How to Improve Research Practices}.
\newblock \bibinfo{journal}{{\em J.\ of Syst.\ \& Softw.\/}}
  \bibinfo{volume}{116} (\bibinfo{year}{2016}), \bibinfo{pages}{133--145}.
\newblock


\bibitem[\protect\citeauthoryear{J{\o}rgensen, Indahl, and
  Sj{\o}berg}{J{\o}rgensen et~al\mbox{.}}{2003}]%
        {Jorg03}
\bibfield{author}{\bibinfo{person}{M. J{\o}rgensen}, \bibinfo{person}{U.
  Indahl}, {and} \bibinfo{person}{D. Sj{\o}berg}.}
  \bibinfo{year}{2003}\natexlab{}.
\newblock \showarticletitle{Software effort estimation by analogy and
  `regression toward the mean'}.
\newblock \bibinfo{journal}{{\em J.\ of Syst.\ \& Softw.\/}}
  \bibinfo{volume}{68}, \bibinfo{number}{3} (\bibinfo{year}{2003}),
  \bibinfo{pages}{253--262}.
\newblock


\bibitem[\protect\citeauthoryear{Kampenes, Dyb{\aa}, Hannay, and
  Sj{\o}berg}{Kampenes et~al\mbox{.}}{2007}]%
        {Kamp07}
\bibfield{author}{\bibinfo{person}{V. Kampenes}, \bibinfo{person}{T. Dyb{\aa}},
  \bibinfo{person}{J. Hannay}, {and} \bibinfo{person}{D. Sj{\o}berg}.}
  \bibinfo{year}{2007}\natexlab{}.
\newblock \showarticletitle{A systematic review of effect size in software
  engineering experiments}.
\newblock \bibinfo{journal}{{\em Information and Software Technology\/}}
  \bibinfo{volume}{49} (\bibinfo{year}{2007}), \bibinfo{pages}{1073--1086}.
\newblock


\bibitem[\protect\citeauthoryear{Kitchenham}{Kitchenham}{2008}]%
        {Kitc08}
\bibfield{author}{\bibinfo{person}{B. Kitchenham}.}
  \bibinfo{year}{2008}\natexlab{}.
\newblock \showarticletitle{The role of replications in empirical software
  engineering --- a word of warning}.
\newblock \bibinfo{journal}{{\em Emp.\ Softw.\ Eng.\/}} \bibinfo{volume}{13},
  \bibinfo{number}{2} (\bibinfo{year}{2008}), \bibinfo{pages}{219--221}.
\newblock


\bibitem[\protect\citeauthoryear{Kitchenham, Dyb{\aa}, and
  J{\o}rgensen}{Kitchenham et~al\mbox{.}}{2004}]%
        {Kitc04}
\bibfield{author}{\bibinfo{person}{B. Kitchenham}, \bibinfo{person}{T.
  Dyb{\aa}}, {and} \bibinfo{person}{M. J{\o}rgensen}.}
  \bibinfo{year}{2004}\natexlab{}.
\newblock \showarticletitle{Evidence-based Software Engineering}. In
  \bibinfo{booktitle}{{\em 26th IEEE International Conference on Software
  Engineering (ICSE 2004)}}. \bibinfo{publisher}{IEEE Computer Society},
  \bibinfo{pages}{273--281}.
\newblock


\bibitem[\protect\citeauthoryear{Lipsey and Wilson}{Lipsey and Wilson}{2001}]%
        {Lips01}
\bibfield{author}{\bibinfo{person}{M. Lipsey} {and} \bibinfo{person}{D.
  Wilson}.} \bibinfo{year}{2001}\natexlab{}.
\newblock \bibinfo{booktitle}{{\em Practical meta-analysis}}.
\newblock \bibinfo{publisher}{Sage Publications}.
\newblock


\bibitem[\protect\citeauthoryear{Loken and Gelman}{Loken and Gelman}{2017}]%
        {Loke17}
\bibfield{author}{\bibinfo{person}{E. Loken} {and} \bibinfo{person}{A.
  Gelman}.} \bibinfo{year}{2017}\natexlab{}.
\newblock \showarticletitle{Measurement error and the replication crisis}.
\newblock \bibinfo{journal}{{\em Science\/}} \bibinfo{volume}{355},
  \bibinfo{number}{6325} (\bibinfo{year}{2017}), \bibinfo{pages}{584--585}.
\newblock


\bibitem[\protect\citeauthoryear{Masicampo and Lalande}{Masicampo and
  Lalande}{2012}]%
        {Masi12}
\bibfield{author}{\bibinfo{person}{E. Masicampo} {and} \bibinfo{person}{D.
  Lalande}.} \bibinfo{year}{2012}\natexlab{}.
\newblock \showarticletitle{A peculiar prevalence of $p$ values just below
  .05}.
\newblock \bibinfo{journal}{{\em Journal of Experimental Psychology\/}}
  \bibinfo{volume}{65}, \bibinfo{number}{11} (\bibinfo{year}{2012}),
  \bibinfo{pages}{2271--2279}.
\newblock


\bibitem[\protect\citeauthoryear{Myrtveit and Stensrud}{Myrtveit and
  Stensrud}{1999}]%
        {Myrt99}
\bibfield{author}{\bibinfo{person}{I. Myrtveit} {and} \bibinfo{person}{E.
  Stensrud}.} \bibinfo{year}{1999}\natexlab{}.
\newblock \showarticletitle{A controlled experiment to assess the benefits of
  estimating with analogy and regression models}.
\newblock \bibinfo{journal}{{\em IEEE Transactions on Software Engineering\/}}
  \bibinfo{volume}{25}, \bibinfo{number}{4} (\bibinfo{year}{1999}),
  \bibinfo{pages}{510--525}.
\newblock


\bibitem[\protect\citeauthoryear{{Open Science Collaboration}}{{Open Science
  Collaboration}}{2015}]%
        {Open15}
\bibfield{author}{\bibinfo{person}{{Open Science Collaboration}}.}
  \bibinfo{year}{2015}\natexlab{}.
\newblock \showarticletitle{Estimating the reproducibility of psychological
  science}.
\newblock \bibinfo{journal}{{\em Science\/}} \bibinfo{volume}{349},
  \bibinfo{number}{6251} (\bibinfo{year}{2015}), \bibinfo{pages}{aac4716--3}.
\newblock


\bibitem[\protect\citeauthoryear{Peng}{Peng}{2011}]%
        {Peng11}
\bibfield{author}{\bibinfo{person}{R. Peng}.} \bibinfo{year}{2011}\natexlab{}.
\newblock \showarticletitle{Reproducible Research in Computational Science}.
\newblock \bibinfo{journal}{{\em Science\/}} \bibinfo{volume}{334},
  \bibinfo{number}{6060} (\bibinfo{year}{2011}), \bibinfo{pages}{1226--1227}.
\newblock


\bibitem[\protect\citeauthoryear{Rosenthal}{Rosenthal}{1979}]%
        {Rose79}
\bibfield{author}{\bibinfo{person}{R. Rosenthal}.}
  \bibinfo{year}{1979}\natexlab{}.
\newblock \showarticletitle{The ``File Drawer Problem" and tolerance for null
  results}.
\newblock \bibinfo{journal}{{\em Psychological Bulletin\/}}
  \bibinfo{volume}{86}, \bibinfo{number}{3} (\bibinfo{year}{1979}),
  \bibinfo{pages}{638--641}.
\newblock


\bibitem[\protect\citeauthoryear{Schmidt and Hunter}{Schmidt and
  Hunter}{2014}]%
        {Schm14}
\bibfield{author}{\bibinfo{person}{F. Schmidt} {and} \bibinfo{person}{J.
  Hunter}.} \bibinfo{year}{2014}\natexlab{}.
\newblock \bibinfo{booktitle}{{\em Methods of meta-analysis: Correcting error
  and bias in research findings}}.
\newblock \bibinfo{publisher}{Sage Publications}.
\newblock


\bibitem[\protect\citeauthoryear{Shepperd and Cartwright}{Shepperd and
  Cartwright}{2005}]%
        {Shep05}
\bibfield{author}{\bibinfo{person}{M. Shepperd} {and} \bibinfo{person}{M.
  Cartwright}.} \bibinfo{year}{2005}\natexlab{}.
\newblock \showarticletitle{A Replication of the Use of Regression Towards the
  Mean ({R2M}) as an Adjustment to Effort Estimation Models}. In
  \bibinfo{booktitle}{{\em 11th IEEE Intl.\ Softw.\ Metrics Symposium
  (Metrics05)}}. \bibinfo{publisher}{Computer Society Press}.
\newblock


\bibitem[\protect\citeauthoryear{Shepperd and Schofield}{Shepperd and
  Schofield}{1997}]%
        {Shep97}
\bibfield{author}{\bibinfo{person}{M. Shepperd} {and} \bibinfo{person}{C.
  Schofield}.} \bibinfo{year}{1997}\natexlab{}.
\newblock \showarticletitle{Estimating software project effort using
  analogies}.
\newblock \bibinfo{journal}{{\em IEEE Transactions on Software Engineering\/}}
  \bibinfo{volume}{23}, \bibinfo{number}{11} (\bibinfo{year}{1997}),
  \bibinfo{pages}{736--743}.
\newblock


\bibitem[\protect\citeauthoryear{Shull, Carver, Vegas, and Juristo}{Shull
  et~al\mbox{.}}{2008}]%
        {Shul08}
\bibfield{author}{\bibinfo{person}{F. Shull}, \bibinfo{person}{J. Carver},
  \bibinfo{person}{S. Vegas}, {and} \bibinfo{person}{N. Juristo}.}
  \bibinfo{year}{2008}\natexlab{}.
\newblock \showarticletitle{The role of replications in empirical software
  engineering}.
\newblock \bibinfo{journal}{{\em Emp.\ Softw.\ Eng.\/}} \bibinfo{volume}{13},
  \bibinfo{number}{2} (\bibinfo{year}{2008}), \bibinfo{pages}{211--218}.
\newblock


\bibitem[\protect\citeauthoryear{Simmons, Nelson, and Simonsohn}{Simmons
  et~al\mbox{.}}{2011}]%
        {Simm11}
\bibfield{author}{\bibinfo{person}{J. Simmons}, \bibinfo{person}{L. Nelson},
  {and} \bibinfo{person}{U. Simonsohn}.} \bibinfo{year}{2011}\natexlab{}.
\newblock \showarticletitle{False-positive psychology: Undisclosed flexibility
  in data collection and analysis allows presenting anything as significant}.
\newblock \bibinfo{journal}{{\em Psychological Science\/}}
  \bibinfo{volume}{22}, \bibinfo{number}{11} (\bibinfo{year}{2011}),
  \bibinfo{pages}{1359--1366}.
\newblock


\bibitem[\protect\citeauthoryear{Spence and Stanley}{Spence and
  Stanley}{2016}]%
        {Spen16}
\bibfield{author}{\bibinfo{person}{J. Spence} {and} \bibinfo{person}{D.
  Stanley}.} \bibinfo{year}{2016}\natexlab{}.
\newblock \showarticletitle{Prediction Interval: What to Expect When You're
  Expecting \textellipsis~A Replication}.
\newblock \bibinfo{journal}{{\em PloS {ONE}\/}} \bibinfo{volume}{11},
  \bibinfo{number}{9} (\bibinfo{year}{2016}), \bibinfo{pages}{e0162874}.
\newblock


\bibitem[\protect\citeauthoryear{Stanley and Spence}{Stanley and
  Spence}{2014}]%
        {Stan14}
\bibfield{author}{\bibinfo{person}{D. Stanley} {and} \bibinfo{person}{J.
  Spence}.} \bibinfo{year}{2014}\natexlab{}.
\newblock \showarticletitle{Expectations for replications are yours realistic?}
\newblock \bibinfo{journal}{{\em Perspectives on Psychological Science\/}}
  \bibinfo{volume}{9}, \bibinfo{number}{3} (\bibinfo{year}{2014}),
  \bibinfo{pages}{305--318}.
\newblock


\bibitem[\protect\citeauthoryear{Viechtbauer}{Viechtbauer}{2010}]%
        {Viec10}
\bibfield{author}{\bibinfo{person}{W. Viechtbauer}.}
  \bibinfo{year}{2010}\natexlab{}.
\newblock \showarticletitle{Conducting meta-analyses in R with the metafor
  package}.
\newblock \bibinfo{journal}{{\em Journal of Statistical Software\/}}
  \bibinfo{volume}{36}, \bibinfo{number}{3} (\bibinfo{year}{2010}),
  \bibinfo{pages}{1--48}.
\newblock


\bibitem[\protect\citeauthoryear{Wilcox}{Wilcox}{2012}]%
        {Wilc12}
\bibfield{author}{\bibinfo{person}{R. Wilcox}.}
  \bibinfo{year}{2012}\natexlab{}.
\newblock \bibinfo{booktitle}{{\em Introduction to Robust Estimation and
  Hypothesis Testing\/} (\bibinfo{edition}{3rd} ed.)}.
\newblock \bibinfo{publisher}{Academic Press}.
\newblock


\end{thebibliography}

\end{document}